\documentclass[11pt,a4paper]{article}
\usepackage{jheppub}
\usepackage{graphicx}
\usepackage{bm}% bold math
\def\Vec#1{\mbox{\boldmath $#1$}}

\def\itmb{\begin{itemize}}
\def\itme{\end{itemize}}
\def\enmb{\begin{enumerate}}
\def\enme{\end{enumerate}}
\def\eqnb{\begin{equation}}
\def\eqne{\end{equation}}

\def\PRD{{Phys. Rev.} D}

\begin{document}

\title{Cartan's Supersymmetry and the Decay of a $H^0(0^+)$}  
%\classification{12.38.Aw,12.38.Gc,12.90,11.15.Ha}


%\titlerunning{nparticle and triality}        % if too long for running head
\author{Sadataka Furui }
\affiliation{
 Graduate School of Science and Engineering, Teikyo University\\
2-17-12 Toyosatodai, Utsunomiya, 320-0003 Japan}
\emailAdd{ furui@umb.teikyo-u.ac.jp}

%\author{<author3>}{
%  address={<common address for author2 and author3>}
%  ,altaddress={<author1 address>} % additional visiting address

\abstract{
We compare the decay of a Higgs boson $H^0(0^+)\to\ell\bar\ell \ell\bar\ell\to\gamma\gamma$, $H^0(0^+)\to W\bar W\to \ell\bar\nu\bar\ell\nu$, and $H^0(0^+)\to Z\bar Z\to\ell\bar\ell\ell\bar\ell$ using Cartan's supersymmetry, that defines coupling of a vector particle $x_i$ and Dirac spinors $\psi$ and $\phi$. 

 Apparent discrepancy of $H^0(0^+)\to\gamma\gamma/\ell\bar\ell \ell\bar\ell$ experimental data of  ATLAS and CMS collaborations disappears, since the sum of transition to $\ell\bar\ell \ell\bar\ell$ and $\gamma\gamma$ have the meaning. The trilinearity of leptons vector fields coupling and the
initial wave functions to be large components define Higgs-leptons couplings.
 
The ratio of Higgs boson decay branching ratios $H\to WW=21.6\pm 0.9$, $H\to\gamma\gamma=0.228\pm 0.011$ can be reproduced when we multiply coupling constant $\alpha_s\simeq 1$ for $WW$ and $\alpha_e=1/137$ for $\gamma\gamma$  to the number of independent diagrams. 

Comparison with Atiyah-Witten's quark-gluon dynamics on a manifold of $G_2$ holonomy theory is added.
}
%%\begin{document}
\maketitle
\section{Introduction}
The decay properties of Higgs boson $H^0(0^+)$ of mass 125GeV is not well understood. It decays as 
$H\to WW=21.6\pm 0.9\%$, $H\to\gamma\gamma=0.228\pm 0.011\%$ and $H\to ZZ=2.67\pm0.11\%$\cite{ATLAS_CMS15a}. 

The Higgs boson decay to $\gamma\gamma$ is not simple, since $Z\bar Z$ decays into $\ell\bar\ell \ell\bar\ell$ and $\ell\bar\ell$ can make a $\gamma$. The measurements of the decay branching ratios of $H^0\to\gamma\gamma$ and $H^0\to\ell\bar\ell \ell\bar\ell$ by the ATLAS Collaboration and by the CMS collaboration 
\cite{ATLAS13,ATLAS14,CMS14a,CMS14b,CMS14c,CMS15,ATLAS15,ATLAS_CMS15} are not in agreement, but the sum of the branching ratios of $\gamma\gamma$ and $\ell\bar\ell\ell\bar\ell$ are in agreement. 

Experimentally, the ratio of the branching ratio of $\gamma\gamma$ and $W\bar W$ or $Z\bar Z$ are also measured.
The boson $W$ decays in the average 10.86\% to $\ell\bar\nu$, and $Z$ decays in the average 3.6\% to $\ell\bar\ell$ ($\ell=e,\mu$ or $\tau$). The signal strength of $H^0\to W\bar W$ of the ATLAS Collaboration and the CMS collaboration differ by about 20\%\cite{ATLAS_CMS15}.
It is therefore interesting to calculate the decay branching ratios of the Higgs boson to $\gamma$ and vector bosons in a model beyond the standard model. 

In the book of the theory of spinors\cite{Cartan66}, Cartan proposed two types of Dirac spinors $\phi$ and $\psi$ and their charge conjugates ${\mathcal C}\phi$ and ${\mathcal C}\psi$ and two types of vector particles $\Vec E$ and ${\Vec E}'$. 
In \cite{SF14} we studied decays of $H^0(0^+)$ into photons, and in \cite{SF15} we studied the decay of $\chi_b(n\,P)$ into $\Upsilon(m\,S)\gamma$ and possible decay of Higgs partner $h^0(0^+)$ into $\Upsilon(m\,S)\gamma$ ($m=1,2$).  Experimental results of $h^0(0^+)$ are not clear, and in this paper, we study decays of $H^0(0^+)$ into $\ell\bar\ell \ell\bar\ell$ and compare with the decay of $H^0(0^+)\to \ell\bar\ell \ell\bar\ell\to\gamma\gamma$, $H^0(0^+)\to Z\bar Z$ and $H^0(0^+)\to W\bar W$.

The experimental decay branching ratio of $H^0(0^+)\to Z\bar Z$ is obtained from $H^0(0^+)\to \ell\bar\ell \ell\bar\ell$, and the strength parameter in this channel seems to be very sensitive to the mass of $H^0(0^+)$, and whether the final $\ell\bar\ell \ell\bar\ell$ originates from $Z\bar Z$ is not clear. By inclusion of $g\bar g\to \ell\bar\ell \ell\bar\ell$, the signal strength of $H^0(0^+)\to Z\bar Z$ was reduced to 0.99\cite{Gainer14}, and
by inclusion of $g\bar g\to Z\bar Z$, the signal strength was reduced to 0.93\cite{Finco14}.

In the supersymmetric theory, the coupling of the Higgs boson to two leptons is given by
the right chiral leptons, defined as $\mathcal E_i$ ($i=1,2,3)$
\begin{eqnarray}
{\mathcal E}_1&=&\tilde \phi_{\bar e}+\theta\cdot\chi_{\bar e}+\frac{1}{2}\theta\cdot\theta F_{\bar e}\nonumber\\
{\mathcal E}_2&=&\tilde \phi_{\bar \mu}+\theta\cdot\chi_{\bar \mu}+\frac{1}{2}\theta\cdot\theta F_{\bar \mu}\nonumber\\
{\mathcal E}_3&=&\tilde \phi_{\bar \tau}+\theta\cdot\chi_{\bar \tau}+\frac{1}{2}\theta\cdot\theta F_{\bar \tau}\nonumber
\end{eqnarray}
and left chiral leptons, defined as
\begin{eqnarray}
{\mathcal L}_1&=&\left(\begin{array}{c}\tilde \phi_{\nu_e}\\
                                                    \tilde\phi_{e}\end{array}\right)
+\theta\cdot\left(\begin{array}{c}\chi_{\nu_e}\\
                                             \chi_e\end{array}\right)
+\frac{1}{2}\theta\cdot\theta \left(\begin{array}{c} F_{\nu_e}\\
                                                                     F_e\end{array}\right)
\nonumber
\end{eqnarray}
and ${\mathcal L}_2$ and ${\mathcal L}_3$ defined similarly \cite{Labelle10}.

The Higgs coupling to $Z$ through leptons is given by using $\chi_{\nu_{L_i}}$ part of ${\mathcal L}_j$,
\[
-y_\ell^{ij}{\mathcal E}_i({\mathcal L}_j\circ {\mathcal H}_d) =-y_\ell^{ij} \nu_\ell(\bar \ell_{L_i} \ell_{L_j}+\bar \ell_{L_j} \ell_{L_i})  
\]
where $L_i, L_j$ specify left handed $SU(2)$ lepton degrees of freedom.

The Higgs coupling to $W$ through leptons is given by 
\[
-y_\ell^{ij}{\mathcal E}_i({\mathcal L}_j\circ {\mathcal H}_d) =-y_\ell^{ij} \nu_\ell(\bar \ell_{L_i} \ell_{\nu_j}) 
\]
and that to $\bar W$ is
\[
-y_\ell^{ji}{\mathcal E}_j({\mathcal L}_i\circ {\mathcal H}_d) =-y_\ell^{ji}\nu_\ell(\ell_{L_j} \bar \ell_{\nu_i} ).
\]

Couplings to quarks of $u,c,t$ of Higgs bosons are defined by
\[
y_u^{ij}{\mathcal U}_i({\mathcal Q}_j\circ {\mathcal H}_u)=y_u^{ij}\, \nu_u\,\bar u_{i} u_{j}.
\]
where $\mathcal Q$ defines the colour and $\mathcal U$ defines the flavour 

In sect.2, we formulate Higgs boson decay into $\gamma\gamma$ using  Cartan's supersymmetric theory of spinors, and  in sect.3  Higgs boson decay into vector bosons 
using Cartan's supersymmetry. 
Nature of the $S(750$ GeV) and a possible enhancement of $\gamma\gamma\to S$ at 13GeV 
from Cartan's supersymmetry in dicussed in sect.4.
Discussion and conclusion are given in sect.5. 

\section{Analysis of Higgs boson decay into $\gamma\gamma$ using Cartan's supersymmetry}
    Following Cartan's supersymmetry, we express scalar particles by $\Psi$ and $\Phi$, and leptons by $\psi, {\mathcal C}\phi$ and antileptons by ${\mathcal C}\psi, \phi$\cite{Cartan66}.  
The coupling of a vector field $x_i (i=1,2,3,4)$ to Dirac spinors,
\begin{eqnarray}
\psi&=&\xi_1\Vec i+\xi_2\Vec j+\xi_3\Vec k+\xi_4\Vec I
=\left(\begin{array}{cc}\xi_4+i\xi_3 &i \xi_1-\xi_2\\
                                i\xi_1+\xi_2&\xi_4-i\xi_3\end{array}\right)\nonumber\\
C\psi&=&-\xi_{234}\Vec i-\xi_{314}\Vec j-\xi_{124}\Vec k+\xi_{123}\Vec I
=\left(\begin{array}{cc}\xi_{123}-i\xi_{124}&-i\xi_{234}+\xi_{314}\\
                                -i\xi_{234}-\xi_{314}&\xi_{123}+i\xi_{124}\end{array}\right)
\end{eqnarray}
and the spinor operator
\begin{eqnarray}
\phi&=&\xi_{14}\Vec i+\xi_{24}\Vec j+\xi_{34}\Vec k+\xi_0\Vec I
=\left(\begin{array}{cc}\xi_0+i\xi_{34} &i \xi_{14}-\xi_{24}\\
                                i\xi_{14}+\xi_{24}&\xi_0-i\xi_{34}\end{array}\right)\nonumber\\
C\phi&=&-\xi_{23}\Vec i-\xi_{31}\Vec j-\xi_{12}\Vec k+\xi_{1234}\Vec I
=\left(\begin{array}{cc}\xi_{1234}-i\xi_{12} &-i \xi_{23}+\xi_{31}\\
                                -i\xi_{23}-\xi_{31}&\xi_{1234}+i\xi_{12}\end{array}\right)
\end{eqnarray}
are specified by trilinearity \cite{Cartan66,SF12a,SF12b}
\begin{eqnarray}
{\mathcal F}_H&=&^t\phi CX\psi={^t(}C\phi)\gamma_0 x_{i}\gamma_i\psi+
{^t\phi} \gamma_0 x_{i'}\gamma_i (C\psi)\nonumber\\
&=&x_1(\xi_{12}\xi_{314}-\xi_{31}\xi_{124}-\xi_{14}\xi_{123}+\xi_{1234}\xi_1)\nonumber\\
&+&x_2(\xi_{23}\xi_{124}-\xi_{12}\xi_{234}-\xi_{24}\xi_{123}+\xi_{1234}\xi_2)\nonumber\\
&+&x_3(\xi_{31}\xi_{234}-\xi_{23}\xi_{314}-\xi_{34}\xi_{123}+\xi_{1234}\xi_3)\nonumber\\
&+&x_4(-\xi_{14}\xi_{234}-\xi_{24}\xi_{314}-\xi_{34}\xi_{124}+\xi_{1234}\xi_4)\nonumber\\
&+&x_{1'}(-\xi_{0}\xi_{234}+\xi_{23}\xi_{4}-\xi_{24}\xi_{3}+\xi_{34}\xi_2)\nonumber\\
&+&x_{2'}(-\xi_{0}\xi_{314}+\xi_{31}\xi_{4}-\xi_{34}\xi_{1}+\xi_{14}\xi_3)\nonumber\\
&+&x_{3'}(-\xi_{0}\xi_{124}+\xi_{12}\xi_{4}-\xi_{14}\xi_{2}+\xi_{24}\xi_1)\nonumber\\
&+&x_{4'}(\xi_{0}\xi_{123}-\xi_{23}\xi_{1}-\xi_{31}\xi_{2}-\xi_{12}\xi_3)
\end{eqnarray}

\begin{figure}[htb]
\begin{minipage}[b]{0.47\linewidth}
\begin{center}
\includegraphics[width=4cm,angle=0,clip]{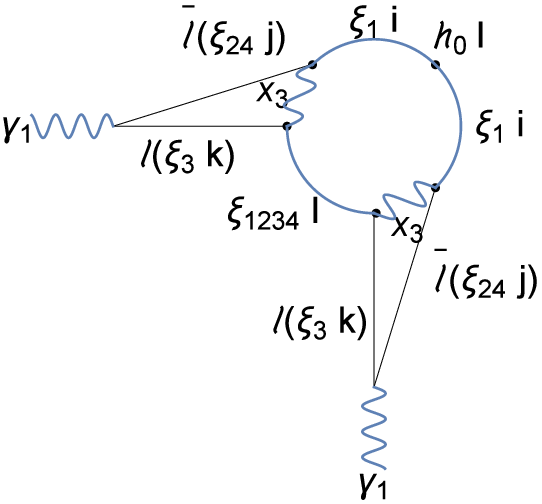}
\end{center}
\end{minipage}
\hfill
\begin{minipage}[b]{0.47\linewidth}
\begin{center}
\includegraphics[width=4cm,angle=0,clip]{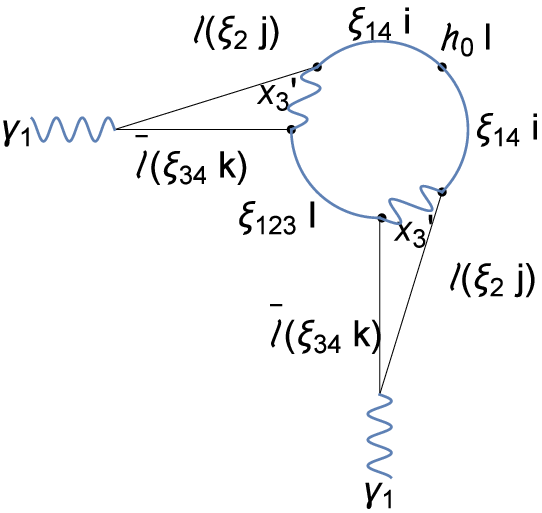}
\end{center}
\end{minipage}
%\end{figure}
%\begin{figure}[htb]
\begin{minipage}[b]{0.47\linewidth}
\begin{center}
\includegraphics[width=4cm,angle=0,clip]{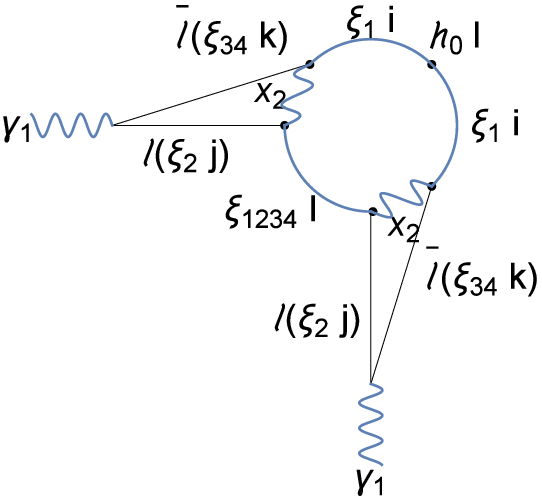}
\end{center}
\end{minipage}
\hfill
\begin{minipage}[b]{0.47\linewidth}
\begin{center}
\includegraphics[width=4cm,angle=0,clip]{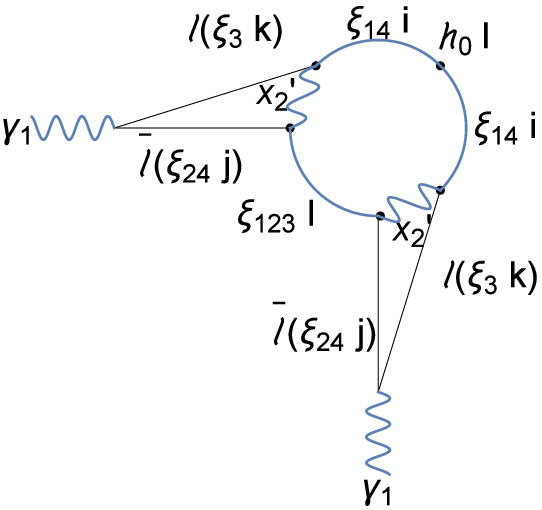}
\end{center}
\end{minipage}
\begin{minipage}[b]{0.47\linewidth}
\begin{center}
\includegraphics[width=4cm,angle=0,clip]{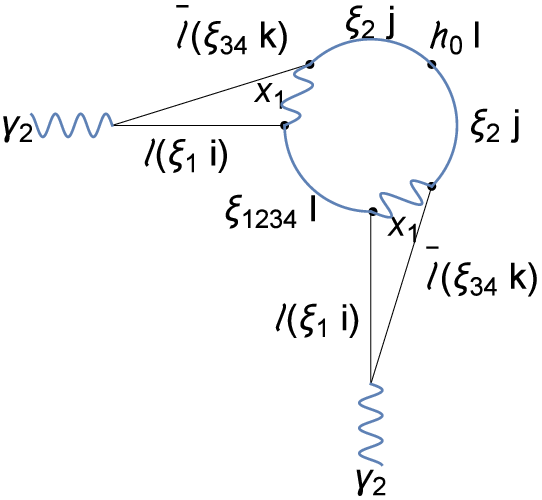}
\end{center}
\end{minipage}
\hfill
\begin{minipage}[b]{0.47\linewidth}
\begin{center}
\includegraphics[width=4cm,angle=0,clip]{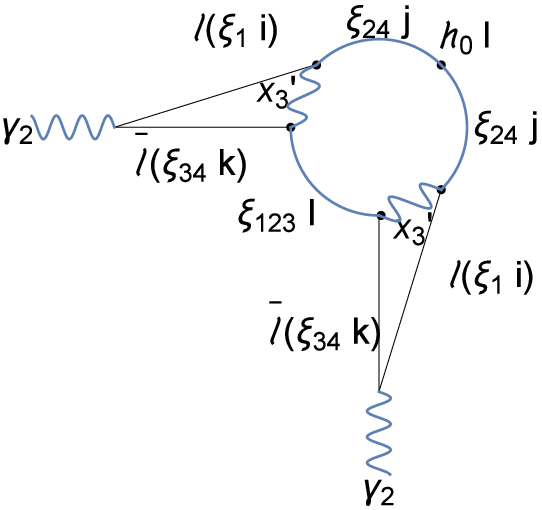}
\end{center}
\end{minipage}
%\end{figure}
%\begin{figure}[htb]
\begin{minipage}[b]{0.47\linewidth}
\begin{center}
\includegraphics[width=4cm,angle=0,clip]{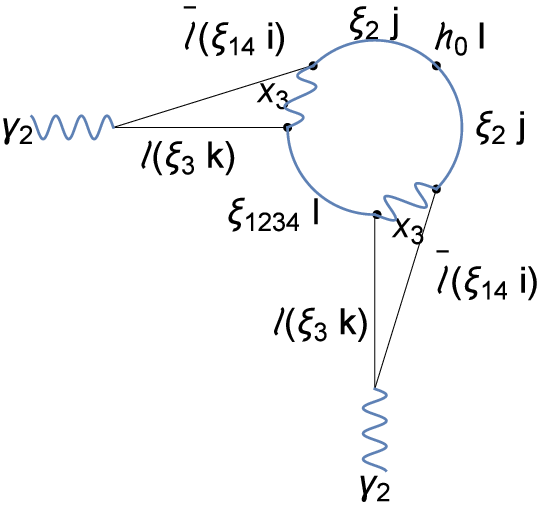}
\end{center}
\end{minipage}
\hfill
\begin{minipage}[b]{0.47\linewidth}
\begin{center}
\includegraphics[width=4cm,angle=0,clip]{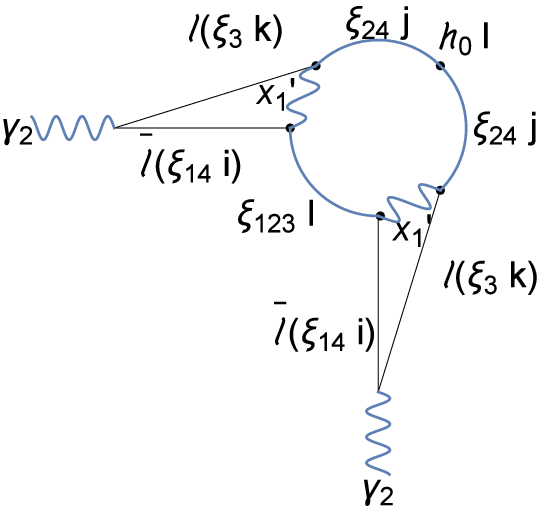}
\end{center}
\end{minipage}
\begin{minipage}[b]{0.47\linewidth}
\begin{center}
\includegraphics[width=4cm,angle=0,clip]{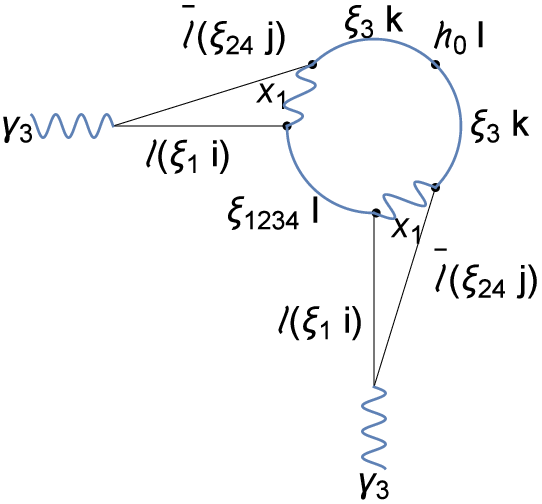}
\end{center}
\end{minipage}
\hfill
\begin{minipage}[b]{0.47\linewidth}
\begin{center}
\includegraphics[width=4cm,angle=0,clip]{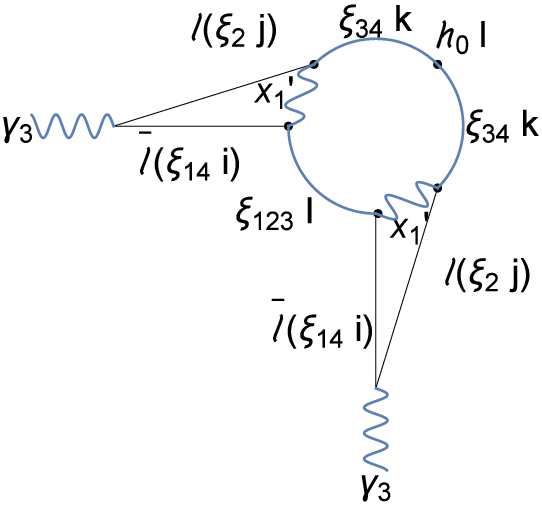}
\end{center}
\end{minipage}
\begin{minipage}[b]{0.47\linewidth}
\begin{center}
\includegraphics[width=4cm,angle=0,clip]{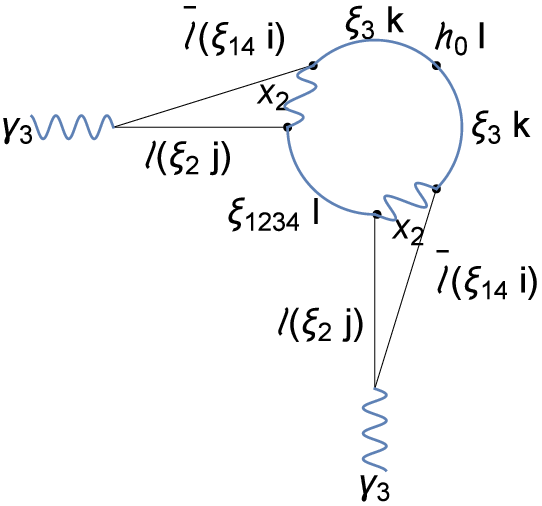}
\end{center}
\end{minipage}
\hfill
\begin{minipage}[b]{0.47\linewidth}
\begin{center}
\includegraphics[width=4cm,angle=0,clip]{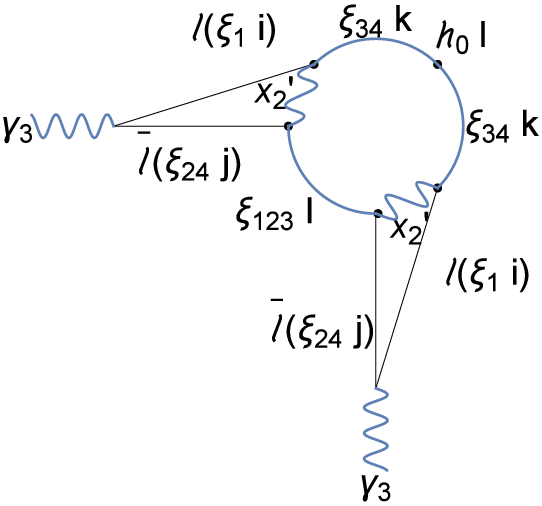}
\end{center}
\end{minipage}
\caption{Higgs boson decay into $\gamma(\ell\bar\ell)\gamma(\ell\bar\ell)$.  }
\label{HiggsA}
\end{figure}

In the case of coupling of $\gamma\gamma$ to a Higgs boson, I  first fix the types of polarization of $\ell\bar\ell$ that creates $\gamma$ which are expressed by large components of Dirac wave functions, and couple quarks of different $q\bar q$ pairs by $\xi_{1234}$ and antiquarks to Higgs boson as shown in the left hand side of  Fig.\ref{HiggsA}. The right hand side of Fig.\ref{HiggsA} are diagrams in which roles of quarks and antiquarks are interchanged and $\xi_{1234}$ is interchanged to $\xi_{123}$.

In these 12 diagrams, lepton pairs which move are components of $(\psi,\mathcal C\psi)$ and $(\phi,\mathcal C \phi)$, which can be regarded as relatively anti-particles and produce a $\gamma$.

Couplings of quarks to electromagnetic field is defined as
\[
{\mathcal F}_L=\bar\psi_L^k (\gamma_L(i\partial_\mu-e A_\mu)-m)\psi_L^k+
          \bar\phi_L^k (\gamma_L(i\partial_\mu-e A_\mu)-m)\phi_L^k
\]
where
\[
\psi_L^k=\left(\begin{array}{c} \psi^k\\
                                       {\mathcal C}\psi^k\end{array}\right) \, {\rm and}\,
\phi_L^k=\left(\begin{array}{c} {\mathcal C}\phi^k \\
                              \phi^k\end{array}\right).                                                     
\]
In electromagnetic interactions $\psi$ v.s. ${\mathcal C}\phi$ and $\phi$ v.s. ${\mathcal C}\psi$ have no differences, but we expect that our electronic detector is sensitive only to $\psi_L^k$.  

When the momenta of the 4 leptons are parallel, $H^0$ decay to $\ell\bar\ell \ell\bar\ell$ and to $\gamma\gamma$ through exchange of vector particle $X$ is possible. However  kinematically the configuration is suppressed, and we expect $H^0$ decay to $\gamma\gamma$ mainly occurs via $\ell\bar\ell \ell\bar\ell$ channels.  

\section{Analysis of Higgs boson decay into vector bosons using Cartan's supersymmetry}

In the standard model, the vector boson $Z$ and the vector field $A$ are expressed as
\begin{eqnarray}
Z_\mu&=&\cos\theta_W\cdot {W^3}_\mu+\sin\theta_W\cdot B_\mu\nonumber\\
A_\mu&=&-\sin \theta_W\cdot {W^3}_\mu+\cos\theta_W \cdot B_\mu\nonumber
\end{eqnarray}
and $\sin^2\theta_W\simeq 0.22$ is the Weinberg angle. Relative strength of $B_\mu$ near the energy of the mass of Higgs boson is not so clear.

The  Lagrangian of $W^{\pm}$ and $Z$ is given by
\[
\mathcal L=m_W^2W^{+\mu}W_\mu^-+\frac{1}{2}m_z^2Z_\mu Z^\mu
\] 
and we define $W_\mu^{\pm}=\frac{1}{\sqrt 2}(W^1_{\mu}\pm i\,W^2_{\mu})$
and 
\[
\frac{1}{2}m_Z^2Z_\mu Z^\mu=\frac{\nu^2}{4}(g W^3-g'B)^2
\]

In contrast to the field of $\gamma$, I make a Cayley-Dickson doubling process and assign complex fields at the position of the Higgs potential. Couplings of quark and antiquarks to $W$ which are expressed as $X_1\pm i\,X_2$ are assumed to satisfy the same trilinearity.
\begin{figure}[htb]
\begin{minipage}[b]{0.47\linewidth}
\begin{center}
\includegraphics[width=4cm,angle=0,clip]{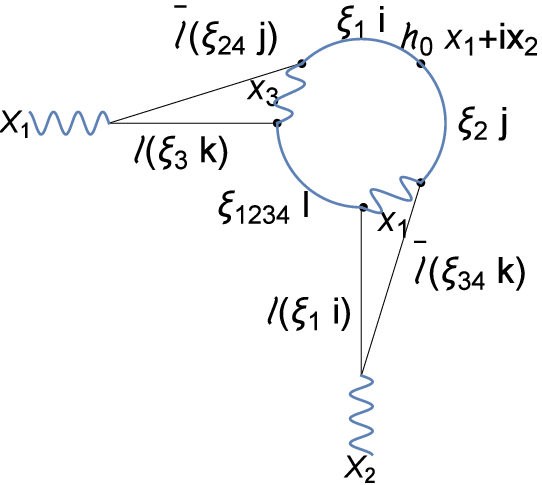}  %
\end{center}
\end{minipage}
\hfill
\begin{minipage}[b]{0.47\linewidth}
\begin{center}
\includegraphics[width=4cm,angle=0,clip]{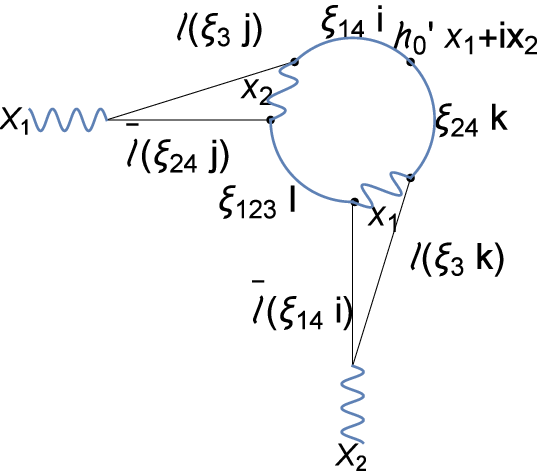}
\end{center}
\end{minipage}
\begin{minipage}[b]{0.47\linewidth}
\begin{center}
\includegraphics[width=4cm,angle=0,clip]{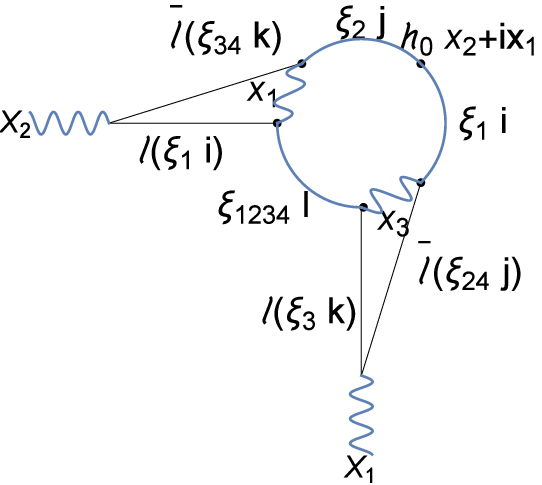}
\end{center}
\end{minipage}
\hfill
\begin{minipage}[b]{0.47\linewidth}
\begin{center}
\includegraphics[width=4cm,angle=0,clip]{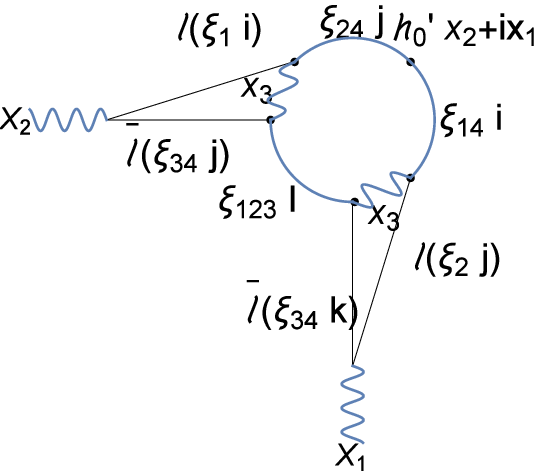}
\end{center}
\end{minipage}
\caption{Higgs boson decay into $W(\ell\bar\nu)W(\bar\ell\nu)$.  }
\label{HiggsWW}
\end{figure}

Couplings of $W^\pm$ fields to the Higgs boson are described by  $(X_1+i\,X_2)(X_1-i\, X_2)=X_1^2+X_2^2$. Since $X_1+i\,X_2$ and$X_1-i\,X_2$ are independent fields 
there are 8 independent diagrams.

\begin{figure}[htb]
\begin{minipage}[b]{0.47\linewidth}
\begin{center}
\includegraphics[width=4cm,angle=0,clip]{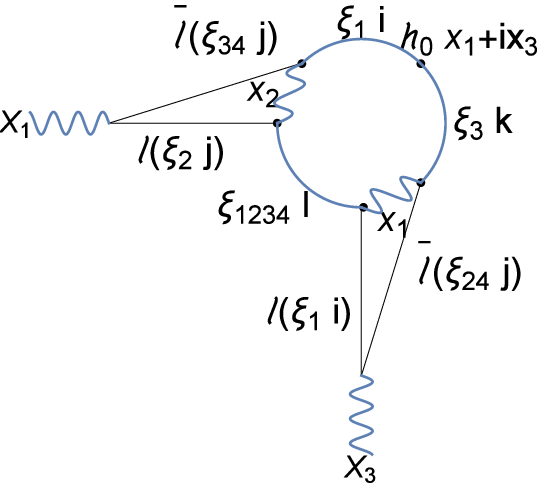}
\end{center}
\end{minipage}
\hfill
\begin{minipage}[b]{0.47\linewidth}
\begin{center}
\includegraphics[width=4cm,angle=0,clip]{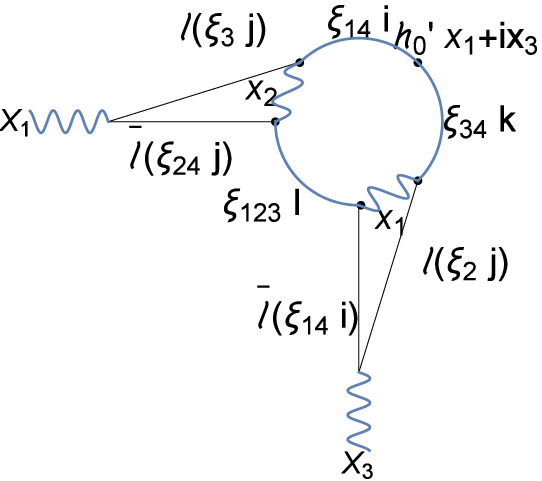}
\end{center}
\end{minipage}
\begin{minipage}[b]{0.47\linewidth}
\begin{center}
\includegraphics[width=4cm,angle=0,clip]{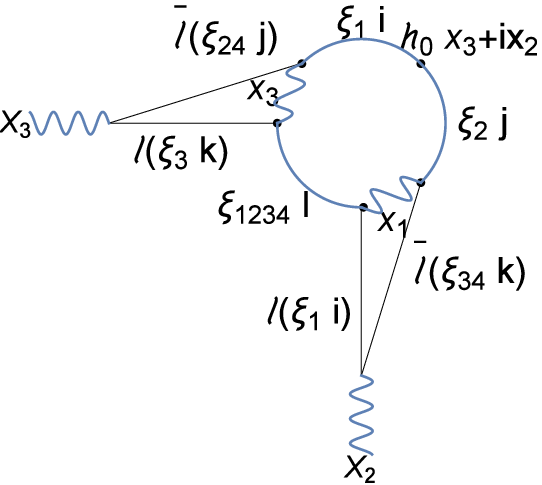}%
\end{center}
\end{minipage}
\hfill
\begin{minipage}[b]{0.47\linewidth}
\begin{center}
\includegraphics[width=4cm,angle=0,clip]{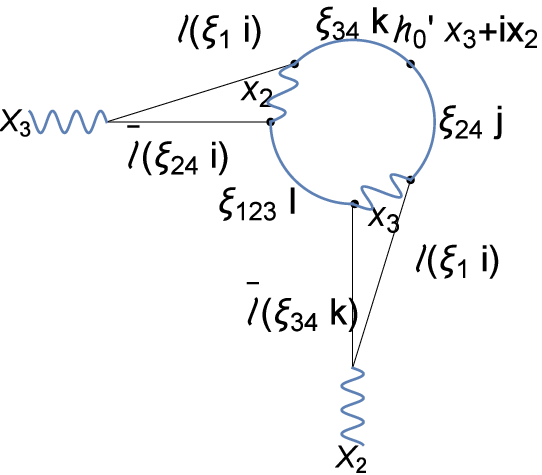}%
\end{center}
\end{minipage}
\caption{Higgs boson decay into $W_3(\ell\bar\nu)B(\bar\ell\ell)$.  }
\label{HiggsWZ}
\end{figure}
In these four diagrams $X_3$ in the final state can be interpreted as $Z(\ell\bar\ell)$. 

In the world of hadron dynamics, photons couple to Higgs boson via $Z$ bosons, and in addition to the coupling to leptons, there are couplings to quarks $d,s,b$ via $y_d^{ij}$ defined by\cite{Labelle10}
\[
-y_d^{ij}{\mathcal D}_i({\mathcal Q}_j\circ {\mathcal H}_d)=-y_d^{ij}\, \nu_d\,\bar d_{i} d_{j}.
\]
Here, $\nu_d$ is defined by the minimum of the potential $H_d={^t(}\nu_d,0)$, and after diagonalization masses of three generations $m_{d,s,b}=\nu_d\, y_{d,s,b}$ appear.

In the weak interactions there are differences in $\phi_L^k$ and $\psi_L^k$,
as the final states of the supersymmetric transformation\cite{SF14}.

In the expression of $-y^{ij}_d \nu_\ell\bar \ell_{L_i}\ell_{L_j}$ coupling,
\[
\ell_{L_j}=\left(\begin{array}{c} \psi^j\\
                                       {\mathcal C}\psi^j\end{array}\right)\,{\rm and}\,
\bar\ell_{L_i}=({\mathcal C}\phi^i, \phi^i).                                                   
\]

In the coupling of $y_\ell^{ij} \mathcal E_i({\mathcal L}_j\circ {\mathcal H}_d)$, $i,j$ specifiy generations of right handed $\bar\ell$ and left-handed $\ell_i$ or $\nu_j$, respectively. The Higgs boson couples to $W$ by $H^0\, \ell\bar\nu \bar\ell\nu$.

The ratio of Higgs boson decay branching ratios $H\to WW=21.6\pm 0.9\%$, $H\to\gamma\gamma=0.228\pm 0.011\%$\cite{ATLAS_CMS15a} can be reproduced when we multiply coupling constant $\alpha_s\simeq 1$ for $WW$ and $\alpha_e=1/137$ for $\gamma\gamma$  to the number of independent diagrams.  Branching ratio $H\to ZZ=2.67\pm0.11\%$ is difficult to interpret, since $Z$ are detectet mainly as $\ell\bar\ell$s, which decay to $\gamma$s. 

The main difference of Higgs decays to $W$, $Z$ and $\gamma$ is that $\gamma$ belongs to $U(1)_{em}$ and vector field $x$ and $x'$ are treated as identical as shown in Fig.\ref{HiggsA} , while $W$ and $Z$ belong to $SU(2)_{w}$ and higgs field $h_0$ and $h_0'$ are treated as different as shown in Fig.\ref{HiggsWW} and \ref{HiggsWZ}.

 In the energy region of  Higgs boson mass, the decay to $\gamma\gamma$ can occur also through $\ell\bar\ell \ell\bar\ell$ in the final state of $Z\bar Z$.  The boson $Z$ has a partner of a vector field $A$ in this energy region. Labelle\cite{Labelle10} assigned the field $A$ as a photon, but we assign the field here as a gluon, and expect the difference of the $H^0$ signal strength of $Z\bar Z$ final state $1.11\pm 0.3$  and that of $W\bar W$ final state $0.87\pm 0.2$ is due to the contamination of  $A=g\bar g\to\ell\bar \ell \ell\bar\ell$ in the final state of $Z\bar Z$.

\section{Nature of the $S(750$ GeV) and an enhancement of $\gamma\gamma\to S$ at 13GeV}
Since leptons and quarks in scalar particles $\Psi$ and $\Phi$ have charge conjugated states, we can extend the same argument starting from the charge conjugated states ${\mathcal C} \Psi$ and ${\mathcal C}\Phi$.

The coherent decays of Higgs pairs on two planes defined by a unique time
\[
(\Psi \, \Phi)^r, \, ({\mathcal C}\Psi\,{\mathcal C}\Phi)^r, \quad 
(\Psi \, \Phi)^g,\, ({\mathcal C}\Psi\,{\mathcal C}\Phi)^g, \quad{\rm and}\quad 
(\Psi\, \Phi)^b,\, ({\mathcal C}\Psi\,{\mathcal C}\Phi)^b 
\]
into $\gamma\gamma$ can make a peak at 125GeV$\times 6=750$GeV scalar meson $S$, which was observed in LHC experiments of run energy $\sqrt s=13$TeV \cite{ATLAS15a,CMS15a}.

Review of electroweak symmetry breaking in Higgs boson decay was given in \cite{Djouadi05}.
In the standard theory, a large component of a lepton or a quark remains large component after coupling to photons. But Cartan's couplings of $\gamma$ to spinors contain $\gamma$ couplings to $\ell\bar\ell$ or $q\bar q$ vertices.

There are discussion on parent states of the $S$ meson and decays of the $S$ meson into dark matter \cite{Franceschini15,SSSU16}. We think dark matters are weakly interacting particles but electromagenetic waves emitted from them are transformed by $G_{12}, G_{13}, G_{123}$ and $G_{132}$ and  insensitive to our detectors.

There are arguments on the decay width of 750 GeV resonance $S$ to $\gamma\gamma$ and the $W_R$ vector of 1.9TeV decay to $W\,Z$ based on trinification model\cite{HS15,PSV15}
\[
\frac{\Gamma({W_R}^\pm\to W^\pm Z)}{\Gamma(S\to \gamma\gamma)}=\frac{(1\pm 13)\,\rm{fb}}{(8\pm 2)\,\rm{ fb}}.
\]
The mass of parent state ${W_R}^\pm\to W^\pm Z$ is about 3 times the mass of $S$, and which is about 1/6 of the energy of the parent state of $S\to\gamma\gamma$= 12TeV. The factor 1/2 can be understood as due to a choice of the charge $+$ or $-$ and 1/3 can be understood as due to a choice of the colour.

\section{Discussion and conclusion}
We studied the decay of Higgs boson $H^0(0^+)$ to $\gamma\gamma$ via $\ell\bar\ell
\ell\bar\ell$ states. Cartan's supersymmetry allows presence of a scalar boson $\Psi$ and a scalar boson $\Phi$ which decay into $\ell\bar\ell \ell\bar\ell$, and reduce to $\gamma\gamma$. It is important that the direct decay of $\Psi$ or $\Phi$ to $\gamma\gamma$ is not allowed. Experimental discrepancy of the branching ratios of $\gamma\gamma$ and $\ell\bar\ell \ell\bar\ell$ of the ATLAS collaboration and the CMS collaboration \cite{ATLAS_CMS15} is not a problem since branching ratio of $\gamma\gamma$ comes from $\ell\bar\ell \ell\bar\ell$ in our model, and the sum of the branching ratios of $\gamma\gamma$ and $\ell\bar\ell \ell\bar\ell$ has the physical meaning. 

The Higgs meson of 125 GeV expects $S$ meson of mass 750 GeV whose hint was observed in $\sqrt s=13$TeV, since coherent sum of 3 colour states times 2 charge conjugates makes $6\times 125=$750GeV. If there is a parent state of $S$ mesons of energy $2^4\times 750$GeV=12TeV, where the factor $2^4$ comes from the selection of $S\bar S$ or $\bar S S$ in 4 dimensional space-time of  $S$ or $\bar S$, we can understand disappearance of $S$ meson in $\sqrt s=7$ and 8 TeV experiment. 

 When the supersymmetric transformation is $G_{13}$, transformations ${\Vec E}\to {\mathcal C}\psi$ and ${\Vec E}'\to \psi$ in the lepton sector, and the spinor $\phi_L^k\to \tilde{\phi_L^k}$ appears in the quark or lepton sector, where $\tilde{\phi_L^k}$ means that particle-antiparticle transformation is done on  $\phi_L^k$\cite{SF15a}.
When the transformation is $G_{12}$, transformations ${\Vec E}\to \tilde\phi$ in the lepton sector and $\psi_L^k\to\tilde\psi_L^k$ appears in the quark or lepton sector.

When the transformation is $G_{132}$, transformations ${\Vec E}\to \phi$ and ${\Vec E}'\to {\mathcal C}\phi$ occurs in the quark or lepton sector and the spinor $\phi_L^k\to \tilde{\psi_L^k}$ appears in the quark or lepton sector.
When the transformation is $G_{123}$, transformations of ${\Vec E}\to \tilde{{\mathcal C}\psi}$ and ${\Vec E}'\to \tilde{\psi}$ occurs in the quark or lepton sector and $\psi_L^k\to\tilde{\phi_L^k}$ in  the quark or lepton sector.

Possible appearance of $\gamma\gamma$ resonance of 750GeV was proposed in trinification models \cite{HS15, PSV15}, in which Higgs bosons $H_1$ and $H_2$ in some representations of $SU(3)_L\otimes SU(3)_R\otimes SU(3)_c$ gauge symmetry were introduced.
Higgs bosons derived from Cartan's supersymmetry contain trilinearity, and we can work in $SU(2)_L\otimes U(1)_Y\otimes SU(3)_c$ gauge framework.

Corresponding to the triality of Cartan, one can consider three $G_2$ manifolds $X_i$ and a flat $C-$ field of the symmetry class $H^3(Y; U(1))$ and $H^3(X_i;U(1))$\cite{AW02}.

Observation of supersymmetry as a relation between fluctuations in the $C-$ field and fluctuations in metric \cite{AW02} agree with each other is interestig, The fact that the trilinear quark antiquark vector field couplings of Cartan yields Higgs $\gamma\gamma$ decay branching ratio and Higgs $WW$ decay branching ratio consistent with experiments indicates that five dimensional spinor $\xi$ like Kaluza-Klein spinor can play important roles for understanding quark gluon dynamics.. 
Cartan's low energy supersymmetry is based on our number system containing octonions and rich in applicability at multiscale levels.

\vskip 1 true cm

{\bf Acknowledgement}
I thank Stan Brodsky (SLAC) for calling my attention to new experimental results on $H^0$ and $S$ meson decays, Naoki Kondo (Teikyo University) for discussions on Clifford Algebra and Serge Dos Santos (INSA Centre Val de Loire) for helpful comments.

\end{document}